# Temperature dependence of the spin relaxation in highly degenerate ZnO thin films

M. C. Prestgard[1,2], G. Siegel[1,2], R. Roundy[2,3], M. Raikh[2,3], and A. Tiwari[1,2,*]

[1]*Department of Materials Science and Engineering, University of Utah, Salt Lake City, Utah 84112, USA*
[2]*Materials Research Science and Engineering Center, University of Utah, Salt Lake City, Utah 84112, USA*
[3]*Department of Physics & Astronomy, University of Utah, Salt Lake City, Utah 84112, USA*

Zinc oxide is a wide-bandgap semiconductor which is considered a potential candidate for fabricating next-generation transparent spintronic devices. However, before this can be practically achieved, a thorough, scientific understanding of the various spin transport and relaxation processes undergone in this material is essential. In the present paper we report our investigations into these processes via temperature dependent, non-local Hanle experiments. Epitaxial ZnO thin films were deposited on c-axis sapphire substrates using a pulsed laser deposition technique. Careful structural, optical, and electrical characterizations of the films were performed. Temperature dependent Hanle measurements were carried out, using an all-electrical scheme for spin injection and detection, in a non-local geometry over the temperature range of 20 – 300 K. Carrier concentration in these films, as determined by Hall effect measurements, was found to be of the order of $10^{19}$ cm$^{-3}$. It was determined that in such a degenerately doped system it is essential to use Fermi-Dirac statistics to explain the transport of carriers in the system. From the Hanle data, spin relaxation time in the ZnO films was determined at different temperatures. Our analysis of the temperature-dependent spin relaxation time data suggests that the dominant mechanism of spin relaxation in ZnO films is the Dyakonov-Perel (DP) mechanism modified for the wurtzite crystalline structure in which a hexagonal c-axis reflection asymmetry is present. As a result of this modification the spin-relaxation rate is linear-in-momentum.

Subject areas: spin transport, semiconductor spintronics, transparent ZnO

[*] Corresponding author, email: tiwari@eng.utah.edu

# I. INTRODUCTION

In current semiconductor device technology research there is a strong emphasis on developing spintronic devices [1, 2]. These are devices that will be capable of utilizing the spin of an electron in place of, or in addition to, its charge, whereas present electronic devices only use the electron's charge [3 - 6]. Spintronic devices are considered essential to compensate for the impending, size-limited failure of Moore's law, as current electronics capabilities will no longer be able to meet future consumer product demands [4, 7]. The field of semiconductor spintronics, despite its evident potential, requires a thorough understanding of spin transport and relaxation processes in semiconducting materials before such devices are ready for application.

Recently there has been a lot of interest in realizing transparent electronics and spintronics. Zinc oxide (ZnO) is considered an ideal material for such applications. It is a transparent semiconductor with a large bandgap of 3.4 eV and a high, room temperature exciton binding energy of 60 meV. A fairly large spin Hall angle has also been recently observed in ZnO films [8]. These observations are quite intriguing because, in ZnO, the valence band splitting is very small (~ 3.5 meV), so in principle the spin-orbit coupling (SOC) in the material should also be small [9]. For example, if we compare the valence band splitting of ZnO with the splitting of GaAs, the former is an almost two order of magnitude smaller than the latter [10]. This difference in splitting should result in a much weaker SOC (and hence much longer spin relaxation times) in ZnO as compared to GaAs. However, recent experiments have shown that this is not the case and imply that our understanding of the various spin transport and relaxation processes going on in ZnO is far from complete [11].

Here we report our experimental and theoretical investigations into the mechanisms of spin transport and relaxation in highly degenerate ZnO thin films. All electrical, four probe

Hanle devices were used to experimentally determine the spin relaxation rate in ZnO via nonlocal Hanle measurements. We attribute the dominant mechanism of the spin relaxation to the linear-in-momentum (k) Dyakonov-Perel (DP) mechanism based on the weak temperature dependence of the spin relaxation time. This mechanism is specific to the wurtzite crystal structure in ZnO [12]. A strong degeneracy of electrons in our samples makes this temperature dependence in relaxation time even weaker.

## II. EXPERIMENTAL METHODS

Epitaxial thin films of ZnO were deposited on c-axis aligned sapphire substrates using a pulsed laser deposition (PLD) technique. For this, a highly dense ceramic target of ZnO was ablated using a pulsed KrF laser (Lambda Physik COMPex Pro Excimer Laser, $\lambda = 248$ nm, pulse width of 25 ns) for 8,000 pulses at a repetition rate of 10 Hz. Depositions were performed under an oxygen pressure of $10^{-4}$ Torr and substrate temperature $700^{o}$C. Ablation by 8,000 pulses gave thin films of approximately 200 nm thickness. The ZnO films thus prepared were characterized using X-ray diffraction (XRD, Philips X'PERT diffractometer), UV-VIS transmittance spectroscopy and temperature dependent (over the temperature range of 20 - 300 K) electrical conductivity and Hall effect measurements. For making the four probe, nonlocal Hanle test devices, a thin barrier layer of MgO (~ 3-5 nm) was first deposited on top of the ZnO using PLD and then a NiFe layer (~ 10 nm) was deposited in a three probe pattern using photolithography and e-beam evaporation techniques (see Fig. 1a). The center NiFe probe was then cut using a focused ion beam (FIB, dual-beam FEI Helios Nanolab 600) to yield two contacts separated by a ~60 nm gap, thereby giving the desired four probe structure (Fig. 1b).

## III. RESULTS AND DISCUSSION

### A. XRD, UV-Vis Spectroscopy, Electrical Conductivity, and Hall Effect Measurements

The XRD results obtained using Cu $K_\alpha$ radiation are shown in Fig. 2a. Only the peaks corresponding to (0002) and (0004) planes of ZnO and the (0006) planes of the sapphire substrate were observed, indicating the highly c-axis aligned growth of the films. Fig. 2b shows the UV-Vis transmission spectroscopy results. These results clearly show that the ZnO thin films were transparent. The inset of Fig. 2b shows the plot of absorption coefficient squared $\alpha^2$, versus energy hv. A linear relationship indicates the direct nature of the film's bandgap [13]. In Fig. 2c, the electrical conductivity vs. temperature data is shown. An increase in the electrical conductivity on increasing the temperature, confirmed the semiconducting behavior of the films.

The Hall effect measurements performed were used to calculate the carrier concentrations, the mobility, and the diffusion coefficient. The carrier concentration is shown as a function of temperature in Fig. 2d. These results showed that there is a slight decrease in carrier concentration with a decrease in temperature and that the carrier concentration is on the order of $10^{19}$ cm$^{-3}$. From the measured values of the carrier concentration we calculated the diffusion coefficient D, in two steps. First, we determined the temperature dependence of the mobility, $\mu = \sigma/en$, from the measured conductivity $\sigma$, and calculated n. The dependence $\mu(T)$ is shown in Fig. 2e.

As a second step, we employ the Einstein relation to find D, which defines D according to the following relation:

$$D = \frac{\mu \cdot k_B T}{e}, \quad (1)$$

where $k_B$ is the Boltzmann constant, T is the temperature, and e is the elementary charge. The results for this calculation are shown in the inset of Fig. 2f. This relation for diffusion coefficient obeys Boltzmann statistics. However, it is important to note that, with the high electron density in our samples, the Einstein relation should take into account the Fermi-Dirac statistics of electrons [14]. Indeed, for the experimentally determined, room temperature charge carrier density $n = 3.45 \times 10^{19}$ cm$^{-3}$ in our films, the Fermi energy given in eqn. (2):

$$\varepsilon_F = \frac{\hbar^2}{2m^*}(3\pi^2 n)^{2/3}, \quad (2)$$

is approximately 115.9 meV [Fermi temperature, $T_F \sim 1345$ K], where we used the value $m^* = 0.3 m_0$ for the effective mass (m*) in ZnO [16]. For such a high $\varepsilon_F$, the proportionality relation between the diffusion coefficient and the mobility takes the form [14]:

$$D = \frac{\mu(T)k_B T}{e} \frac{F_{1/2}(\frac{\varepsilon_F}{k_B T})}{F_{-1/2}(\frac{\varepsilon_F}{k_B T})}, \quad (3)$$

where $F_j(x)$ is the Fermi-Dirac integral defined as

$$F_j(x) = \frac{1}{j!}\int_0^\infty dt \frac{t^j}{e^{(t-x)}+1}, \quad (4)$$

Note that Fermi-Dirac statistics strongly affect the values of the diffusion coefficients calculated from Einstein relation. Indeed, according to the Boltzmann statistics, the change of temperature from 20 K to 300 K in our experiment should be accompanied by a 15-fold increase in D. However, with Fermi-Dirac integrals in eqn. (3) this increase is only ~1.3 times. Our results for the calculated $D(T)$ according to both Boltzmann and Fermi-Dirac statistics are shown in Fig. 2f. For the room temperature we get $D \approx 3.2 \times 10^{-4}$ m$^2$/s. Following these characterizations, nonlocal, four probe Hanle measurements were performed on the ZnO thin films. As shown in Fig. 1c, current was passed from contact 1 to contact 2 and voltage was measured between the contacts 3 and 4.

## B. Hanle and Hall Effect Measurements

The four probe structure was fabricated in order to perform nonlocal measurements. Traditional three probe Hanle devices utilize the same center contact for both the application of current as well as the measurement of voltage. This method of testing can lead to a convoluted signal, as three probe testing introduces an additional voltage drop, resulting from the contact resistance, to the overall signal. In the four probe structure, any signal measured is a direct result of spin transport occurring within the device, as only spin accumulation is probed [15].

Hanle measurements were then performed using the four probe device at temperatures ranging from 20 – 300 K according to the schematic in Fig. 1c. By passing a current between contacts 1 and 2, the spin polarized electrons is injected at contact 2. Electrons accumulate near the second most contact, and then diffuse towards the third contact. This gives a net spin accumulation at the third contact as well. The voltage measured between contacts 3 and 4 is a result of the difference in spin accumulation between two. Figs. 3a and 3b show the Hanle signal at 300, 200, 100, 50, and 20 K for negative (3a) and positive (3b) applied current, all of which showed the same general shape. A general theoretical expression for describing the shape of the Hanle voltage curve has the form [16, 17]:

$$V_S(B) = \frac{A\sqrt{D\tau}}{\sigma(T)} \cdot \exp\left(\frac{-d}{\sqrt{D\tau}}\right) \cdot (1+\omega^2\tau^2)^{-1/4} \cdot \exp\left(\frac{-d}{\sqrt{D\tau}}\left\{\sqrt{\frac{1}{2}(\sqrt{1+\omega^2\tau^2}+1)}-1\right\}\right)$$
$$\cdot \cos\left\{\frac{\tan^{-1}(\omega\tau)}{2} + \frac{d}{\sqrt{D\tau}}\sqrt{\frac{1}{2}(\sqrt{1+\omega^2\tau^2}-1)}\right\}, \quad (5)$$

where ω is the Larmor frequency, $d$ is the distance between the two inner most contacts, and $\tau$ is the spin lifetime.

The Larmor frequency is given by $\omega = 2\pi g \mu_B B / h$, where g is the Lande g-factor, $\mu_B$ is the Bohr magnetron, and B is the applied magnetic field. For long devices with $d$ much bigger than the spin diffusion length $L_{SD} = \sqrt{D\tau}$, eqn. (5) predicts the $V_S(B)$ dependence with damped oscillations in the tails. Since such oscillation are absent in our data, it is more natural to use the small-$d$ asymptote of eqn. (5),

$$V_S(B) = \frac{A\sqrt{D\tau}}{\sigma(T)\sqrt{2}} \frac{\sqrt{\sqrt{1+\omega^2\tau^2}+1}}{\sqrt{1+\omega^2\tau^2}}, \quad (6)$$

The shape predicted by eqn. (6) is universal, in the sense that it depends only on the product $\omega\tau$. After the value $\tau$ was determined from the fit, the corresponding spin diffusion length was calculated with this $\tau$ and was compared with $d$ to test whether the small-$d$ asymptote applies. Realistically, this approximation is applicable for $d$ up to a few $L_{SD}$. The fit of our data in Figs. 3a and 3b to eqn. (6) for the negative and positive applied current appears to be excellent.

The calculated lifetime values are shown in Fig. 4a. We see that the spin lifetime increases slightly with decreasing temperature between 20 and 300 K and has a value of ~133 ps at room temperature. From this value and the value of the diffusion coefficient estimated above we find the spin diffusion length to be ~200 nm at room temperature, which is more than adequate for the industry standard, 22 nm transistor technology currently available. This length is bigger than d = 60 nm, so that the small-$d$ asymptote, eqn. (6) does in fact apply. Additionally, the temperature dependence of the spin diffusion length is weak, as shown in Fig. 4b.

## C. Discussion

A specific of ZnO, as compared, e.g. to GaAs, is that the wurtzite crystal structure gives rise to a term in the DP spin relaxation rate [18], which is linear-in-momentum, k. This term comes from a linear-in-k spin-orbit term in electron Hamiltonian, which is allowed due to c-axis asymmetry in the ZnO crystal structure [15]. In origin, this term is analogous to the linear-in-k Rashba spin-orbit term in the Hamiltonian of a 2D electron [19]. The general expression for the spin relaxation rate, which is commonly used [19], represents the sum of conventional "cubic" and linear-in-k terms:

$$\frac{1}{\tau_{DP}} = \alpha_{DP}^{(3)} \cdot T^3 \cdot \tau_P(T) + \alpha_{DP}^{(1)} \cdot T \cdot \tau_P(T), \quad (7)$$

where $\alpha_{DP}^{(3)}$ and $\alpha_{DP}^{(1)}$ are the material parameters and $\tau_P(T)$ is the momentum relaxation time. We note that both terms are strongly temperature dependent, so that, according to eqn. (7), the time $\tau_{DP}^{-1}$ should grow at least 15 times between 20 K and 300 K. This is not the case for our data.

Concerning other measurements on ZnO, this sharp temperature dependence of $\tau_{DP}$ has been observed by Ghosh *et al.* [9], where it was measured using Faraday rotation, and in the second sample by Althammer *et al.*, from the spin-valve based electrical measurements [20]. For the other samples in the study by Ghosh *et al.,* [9] the low temperature behavior of $\tau_{DP}(T)$ was essentially flat. In order to account for the apparent discrepancy between this flat behavior and eqn. (7), Harmon *et al.* assumed [19] that an additional spin relaxation mechanism, different from DP, was at work. In this regard, we would like to point out that eqn. (7) applies only for a nondegenerate electron gas. Indeed, for the samples in both Ghosh's [9] and Althammer's [20] studies, which showed a steep temperature dependence of $\tau_{DP}(T)$, the concentrations were low, $10^{15}$-$10^{17}$ cm$^{-3}$. Taking proper care of the electron degeneracy allows us to account for the weak temperature dependence of $\tau_{DP}$ without invoking an additional relaxation mechanism.

Upon incorporating the Fermi-Dirac statistics, the two terms in eqn. (7) take the following form:

$$\frac{1}{\tau_{DP}^{(3)}} = \alpha_{DP}^{(3)} \frac{\int_0^\infty d\varepsilon \cdot g(\varepsilon) \cdot \tau_P(\varepsilon) \cdot \varepsilon^3 \cdot \frac{\partial f_0}{\partial \varepsilon}}{\int_0^\infty d\varepsilon \cdot g(\varepsilon) \cdot \frac{\partial f_0}{\partial \varepsilon}}, \quad (8)$$

$$\frac{1}{\tau_{DP}^{(1)}} = \alpha_{DP}^{(1)} \frac{\int_0^\infty d\varepsilon \cdot g(\varepsilon) \cdot \tau_P(\varepsilon) \cdot \varepsilon \cdot \frac{\partial f_0}{\partial \varepsilon}}{\int_0^\infty d\varepsilon \cdot g(\varepsilon) \cdot \frac{\partial f_0}{\partial \varepsilon}}, \quad (9)$$

Where $f_0$ is the Fermi distribution function, and $g(\varepsilon) \propto \varepsilon^{1/2}$ is the density of states. In terms of the Fermi-Dirac integrals defined above, in eqn. (4), these expressions can be cast in the form

$$\frac{1}{\tau_{DP}^{(3)}} = \alpha_{DP}^{(3)} \cdot T^3 \cdot \tau_p \frac{F_{5/2}\left(\frac{\varepsilon_F}{k_B T}\right)}{F_{-1/2}\left(\frac{\varepsilon_F}{k_B T}\right)}, \quad (10)$$

$$\frac{1}{\tau_{DP}^{(1)}} = \alpha_{DP}^{(1)} \cdot T \cdot \tau_p \frac{F_{1/2}\left(\frac{\varepsilon_F}{k_B T}\right)}{F_{-1/2}\left(\frac{\varepsilon_F}{k_B T}\right)}, \quad (11)$$

In eqns. (10), (11) we neglected the energy dependence of $\tau_P$ since the mobility, $\mu = e\tau_P/m^*$, changes weakly between 20 K and 300 K. Plotting eqns. (10) and (11) for $\varepsilon_F = 1345$ K, we find that, even with degeneracy taken into account, the term $\left(\tau_{DP}^{(3)}\right)^{-1}$ maintains the steep temperature dependence, while the term $\left(\tau_{DP}^{(1)}\right)^{-1}$ describes the measured $T$-dependence of the spin relaxation time without additional mechanisms, which makes the linear-in-k term a better fit to the spin relaxation rate data (Fig. 5). We conclude that it is the linear-in-k, spin-orbit term in the Hamiltonian, specific to the wurtzite structure, which is responsible for the spin relaxation in our ZnO samples.

## IV. CONCLUSIONS

In conclusion, we have shown that degenerately doped ZnO has a relatively long spin diffusion length (180 nm). We demonstrated that electron degeneracy plays a key role in weak temperature dependence of the spin relaxation time observed. We also identified that the leading mechanism of spin relaxation is the linear-in-k Dyakonov-Perel mechanism resulting from the specific crystal structure of ZnO. The values of spin relaxation time $\tau \approx 110$ ps in our degenerate samples agree within a factor of 2 with high-temperature values of $\tau$ in nondegenerate sample in the study by Althammer *et al.* [20].

As a final remark, we note that, aside from the DP mechanism, there are two other spin-orbit-related mechanism which lead to the spin relaxation in semiconductor structures. The Bir-Aronov-Pikus mechanism [21] is based on scattering of electrons by holes, and thus it is inefficient in degenerate n-type structures. Concerning the Elliot-Yafet mechanism [21], in nondegenerate case, the numerical estimates made by Harmon *et al.* [19] demonstrate that it is far too weak to account for the experimental spin-relaxation times.

## ACKNOWLEDGEMENTS

The authors would like to thank the U.S. National Science Foundation for research support through Grant# 1121252 (MRSEC) and Grant# 1234338.

Figure Captions

**Figure 1:** Shows the device schematic used for the non-local Hanle testing. (a) Initially, a three-probe device was fabricated, where the base portion (in gray) is the substrate, on top of that is the ZnO (in purple). Atop of that is a thin layer of MgO (used for spin injection, indicated in green, and the NiFe contact pads (in blue). Following fabrication of the three-probe device, the four-probe device was fabricated using a FIB to cut the center contact in half (b). (c) The device testing was performed as described in this figure. A current was applied between the two left contacts and the voltage, which occurs as a result of diffusing spin accumulation between the center contacts, was measured between the two right contacts.

**Figure 2:** ZnO film characterization techniques: (a) X-ray diffraction showing textured ZnO growth atop the sapphire substrate. (b) UV-Vis spectroscopy showing the transparency of the thin films. The inset shows that the material is a direct bandgap thin film by showing that the plot of absorption coefficient squared versus energy is linear, with an intercept equal to the bandgap of the film. (c) Conductivity versus temperature measurements. (d) Carrier concentration as determined via Hall effect testing. (e) The calculated mobility of carriers in the ZnO. (f) The diffusion coefficient as calculated according to Fermi-Dirac and Boltzmann statistics (Boltzmann distribution shown in the inset).

**Figure 3:** The Hanle curve data and fittings to eqn. 7 for the (a) negative applied current sweeps and (b) the positive applied current sweeps.

**Figure 4:** (a) The spin lifetime as a function of temperature for ZnO. The red and black data points indicate the negative and positive current sweeps, respectively. (b) Shows the calculated spin diffusion length as a function of temperature.

**Figure 5:** The fitting of the spin relaxation data (inverse spin lifetime) to eqns. 11 and 12 for $\varepsilon_F$ between 1000 and 1500 K. The portion in the graph indicated in tan is for the fit to eqn. 11, with the upper bound being for $\varepsilon_F = 1000$ K and the lower bound being from $\varepsilon_F = 1500$ K. The same trend is true of the blue data, which is for eqn. 12. Because our Fermi energy was between ~1000 K and 1500 K at all temperatures, these data sets give an approximate range of spin relaxation rate values for the entire temperature range.

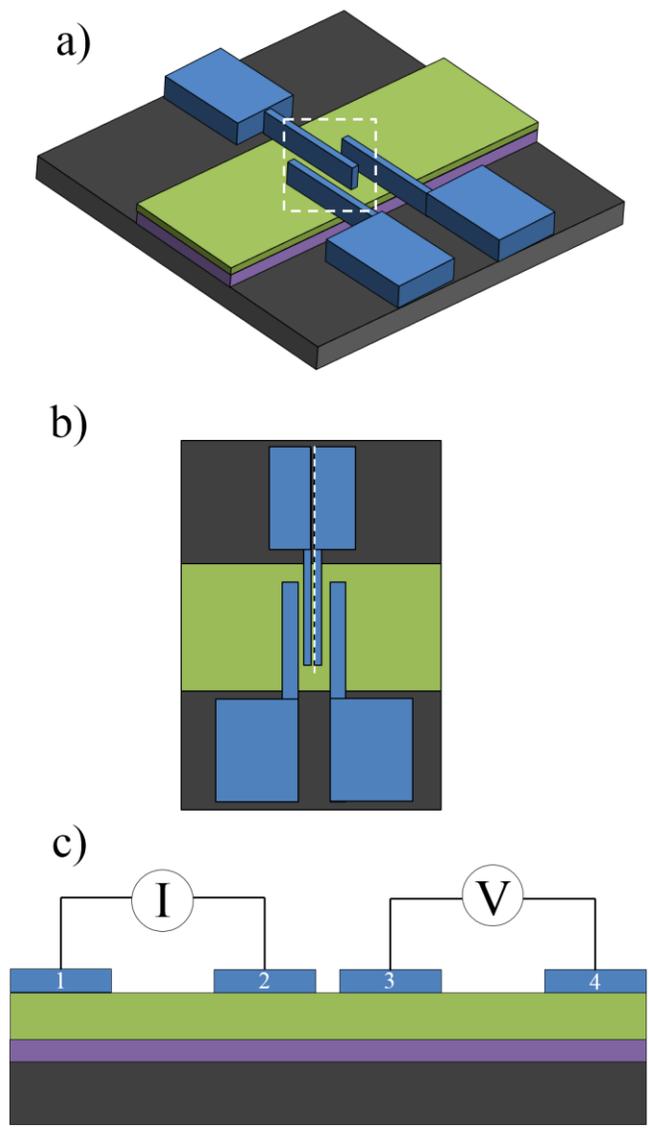

Figure 1

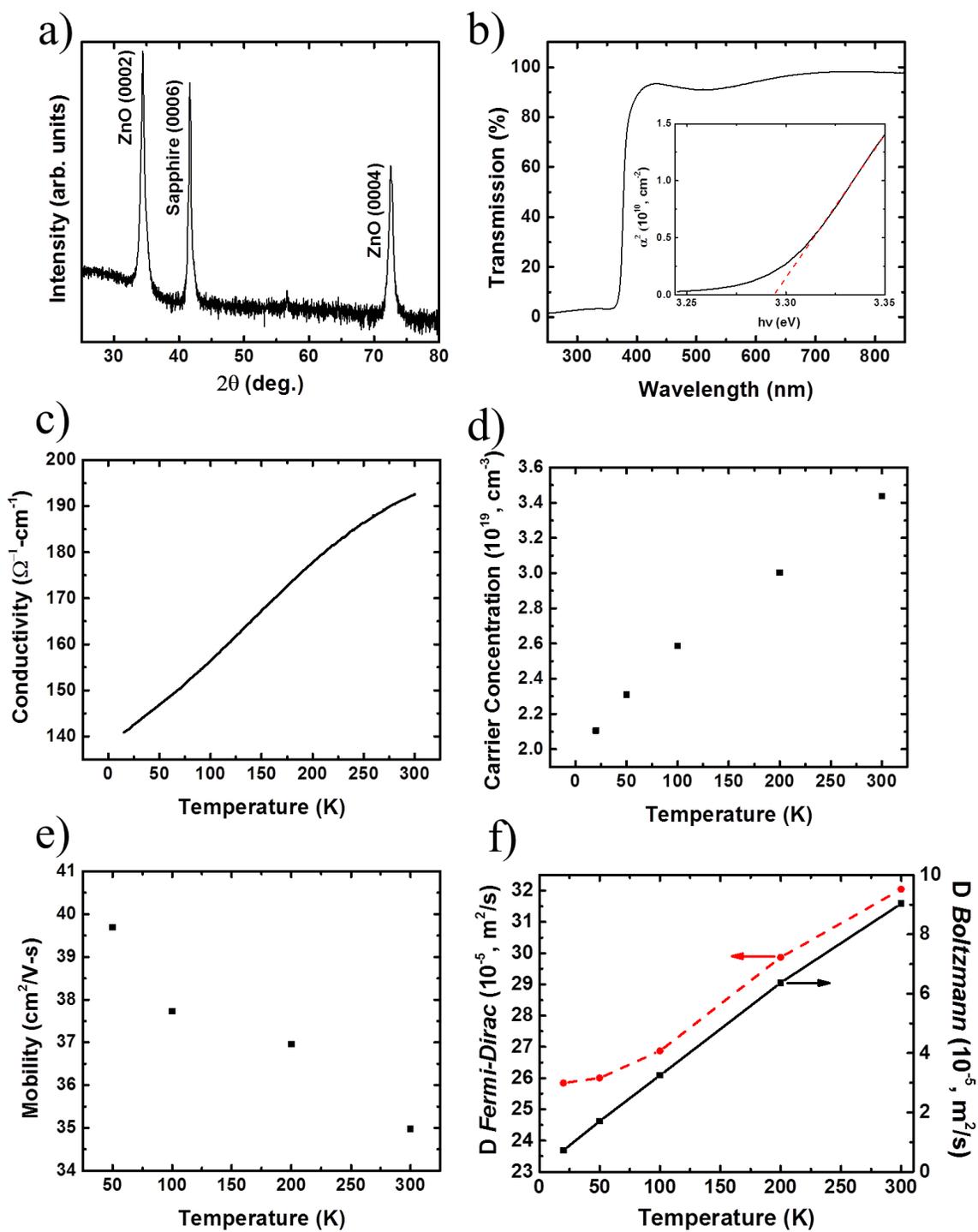

Figure 2

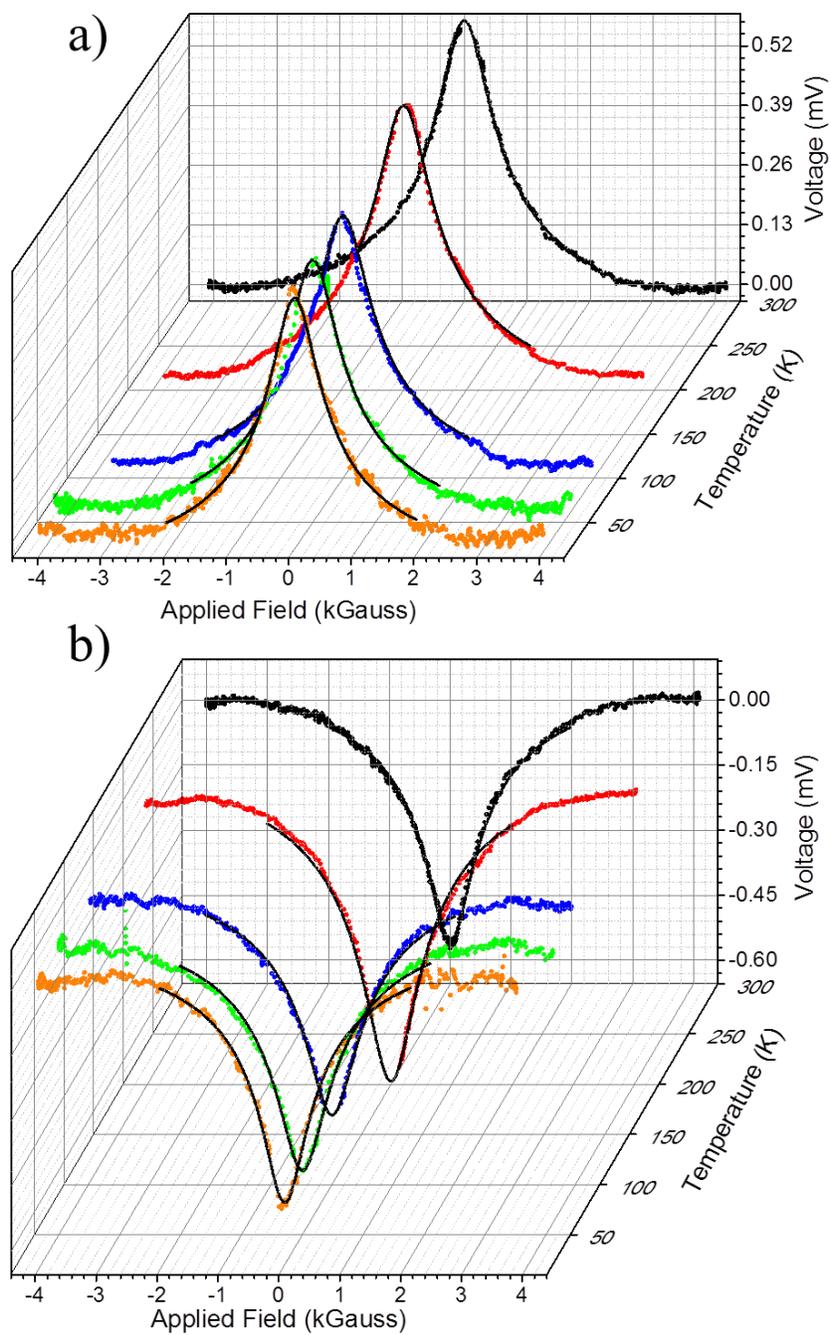

Figure 3

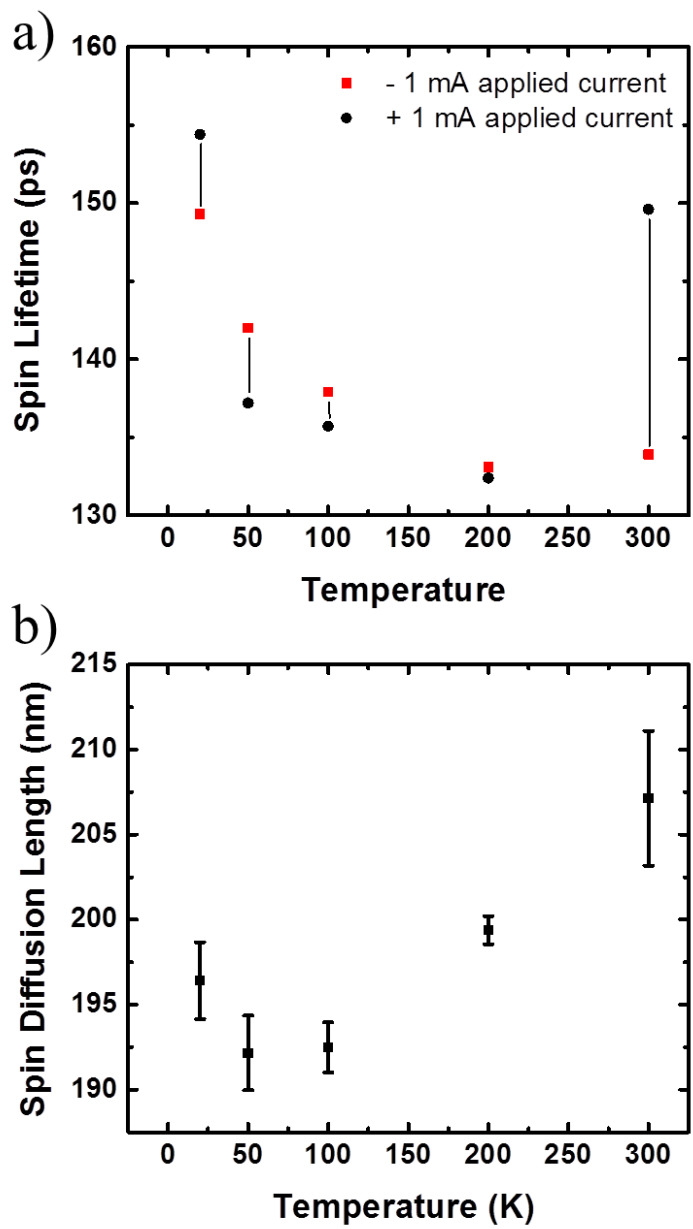

Figure 4

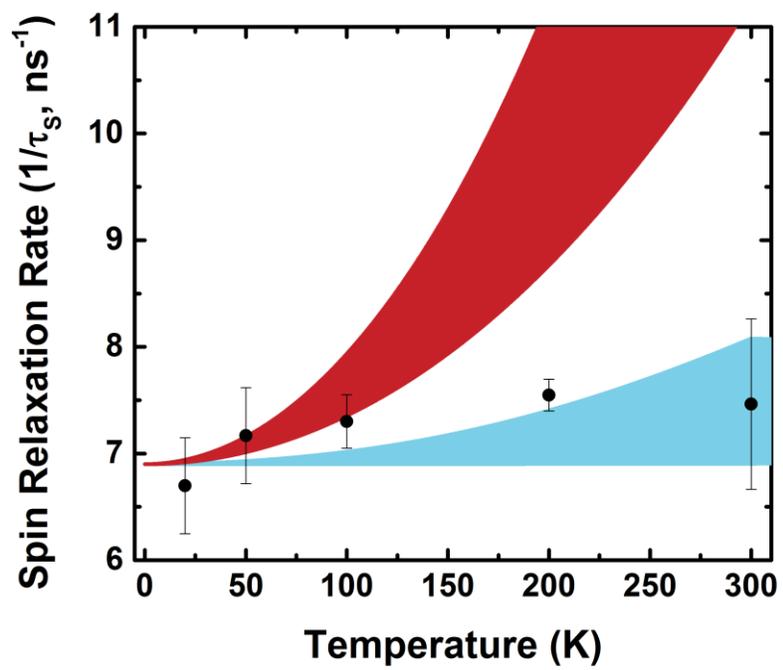

Figure 5